\documentclass[12pt]{iopart}
\usepackage[english]{babel}
\usepackage[utf8]{inputenc}
\usepackage{graphicx}
\usepackage{amssymb}
\usepackage{xcolor}

\newcommand\ten[1]{\mathbf{#1}}

\begin{document}

\title[SSI in stretched soft strips: modeling for quantitative elastography]{Supersonic shear wave imaging in stretched soft strips: modeling for quantitative elastography}
\author{Alexandre Delory, Daniel A. Kiefer, Claire Prada, Fabrice Lemoult}
\address{Institut Langevin, ESPCI Paris, Université PSL, CNRS, 75005 Paris, France}
\ead{fabrice.lemoult@espci.psl.eu}
\vspace{10pt}
\begin{indented}
\item[]December 2023
\end{indented}

\begin{abstract} 
\\\textit{Objective ---} Shear wave elastography has enriched ultrasound medical imaging with quantitative measurements of tissue stiffness.
However, this method still suffers from some limitations due to viscoelasticity, guiding geometry or static deformations.
\\\textit{Approach ---} To explore these limitations, a nearly-incompressible soft elastomer strip is chosen to mimic the mechanical behavior of an elongated tissue. A supersonic shear wave scanner measures the propagation of shear waves within the strip. By repeating the experiment on the same sample for different orientations and static strains, the scanner estimates the shear wave velocity in a wide range from 2 to 6~m/s.
\\\textit{Main results ---} 
The effect of waveguiding is highlighted and the spatio-temporal Fourier transform of the raw data provides dispersion diagrams. We provide a theoretical model that accounts for the static deformation and allows the extraction of the mechanical parameters of the sample, including its rheology and hyperelastic behavior.
\\\textit{Significance ---} To overcome some limitations of current elastography, we propose a method that would allow the simultaneous characterization of the viscoelastic and hyperelastic properties of soft tissues, paving the way for robust quantitative elastography of elongated tissues.
\end{abstract}

\vspace{2pc}
 \vspace{2pc}\\
\maketitle


\section{Introduction}

Elastography is a non-invasive medical imaging technique used to map the elasticity of tissues. It is the modern equivalent of palpation and provides information about the mechanical properties of the tissue by measuring its deformation when subjected to external mechanical forces, such as compression or shear waves. This technique is particularly useful in the field of diagnostic imaging, where it can provide valuable information about the presence and severity of diseases such as liver fibrosis~\cite{sandrin_2003,asbach_2010,deffieux_2015,kennedy_2018}, breast lesions~\cite{sinkus_2005,barr_2012,barr_2019}, prostate cancer~\cite{correas_2013}, thyroid nodules~\cite{cantisani_2015}, heart problems~\cite{elgeti_2014,sinkus_2014,hansen_2015,khan_2018,pruijssen_2020}, tendinopathies~\cite{prado_2018,farron_2009,mifsud_2023} and other musculoskeletal disorders~\cite{winn_2016,paluch_2016,taljanovic_2017,davis_2019}.

Elastography exploits either ultrasound imaging, Magnetic Resonance Imaging (MRI) or Optical Coherence Tomography (OCT) to record a movie from which displacements can be extracted. For each imaging technique, there are several methods for assessing stiffness~\cite{ormachea_2020}.
During the last decade, great progress has been made in Optical Coherence Elastography (OCE) as detailed in comprehensive and recent reviews~\cite{zviet_2022,leartprapun_2023}, and it finds noteworthy applications in ophthalmology~\cite{kirby_2017,lan_2023}. Similarly, Magnetic Resonance Elastography (MRE) has proven its worth~\cite{low_2016,sack_2023} and is also finding clinical applications in large tissues such as breast~\cite{sinkus_2005}, heart~\cite{elgeti_2014,khan_2018,marlevi_2020} and brain~\cite{hiscox_2016}.

Ultrasound imaging, used in this work, is the most widely used method for elastography~\cite{shiina_2014,sigrist_2017} and also finds many clinical applications. In particular, acoustic radiation force methods have proven to be particularly effective to probe the elasticity in real time and in depth~\cite{doherty_2013}.
In this work, Supersonic Shear wave Imaging (SSI), as described in~\cite{bercoff_2004,deffieux_2008}, is used with an Aixplorer\textsuperscript{TM} system. The displacement induced by a push is assessed and the propagation velocity of the shear waves $V_T$ is measured. In an incompressible material of mass density $\rho$, the Young Modulus (YM) can be derived from this velocity using the simple equation $E=3\rho {V_T}^2$.

However, this equation only holds under certain strong assumptions that are rarely valid, thus limiting the robustness of quantitative elastography. These limitations are due to four different reasons. First, the viscoelasticity of a tissue leads to frequency-dependent mechanical parameters, including the derived YM~\cite{sinkus_2014,kennedy_2014,low_2016,hiscox_2016,kirby_2017,kennedy_2018}.
Second, tissues such as muscles are inherently anisotropic and $V_T$ strongly depends on the direction of propagation~\cite{sinkus_2014,prado_2018,mifsud_2023}.
Third, most tissues have boundaries and act as waveguides for shear waves, leading to strong dispersion~\cite{kirby_2017,li_2017,khan_2018,pelivanov_2019,ramier_2019}.
Finally, surrounding fluids or other external factors may apply a prestress on the tissue of interest, again leading to changes in the measured velocity~\cite{elgeti_2014,hansen_2015,cantisani_2015,li_2017,barr_2019}.
It is common that biological tissues combine several of the above aspects as it is highlighted in several reviews using different imaging modalities~\cite{sigrist_2017,bilston_2018,ormachea_2020,caenen_2022,crutison_2022,davis_2019,zviet_2022,leartprapun_2023,lan_2023}. These limitations are known and remain an active area of research.

Viscoelasticity is the most addressed issue~\cite{asbach_2010,gennisson_2010,yasar_2013,brum_2014,deffieux_2015,zampini_2021,juge_2023,sharma_2023} and the waveguiding geometry is also widely studied, especially for clinical applications involving arterial and cardiac walls~\cite{couade_2010,hansen_2015,astaneh_2017,maksuti_2017,caenen_2022}. In addition, several works focus on guided waves in viscoelastic media~\cite{nenadic_2011,nguyen_2011}.
In fact, the body mostly consists of nearly-incompressible and highly deformable media, and the retrieved stiffness depends on the applied stresses~\cite{catheline_2003,gennisson_2007,crutison_2022}.
This dependence is known as the acoustoelastic effect~\cite{biot_1940,toupin_1961} and is not specific to elastography but refers to the changes in elastic wave velocities with a prestress, as described in detail in~\cite{ogden_1997,destrade_2007}. It is also worth mentioning recent works such as~\cite{zhang_2023} where a customized ultrasound sequence was used to consecutively focus pushes along a horizontal line to map stresses in a prestressed soft material by measuring the changes in velocity along two directions. While the guiding geometry, anisotropy and prestress are treated, the viscoelasticity is not accounted for.
A comprehensive review~\cite{li_cao_2017} describes theoretically each of these limits and how they arise in elastography. Of particular interest is the treatment of acoustoelasticity in intrinsically anisotropic media.

Guided elastic waves in a strip are well understood, especially the role of viscoelasticity~\cite{laurent_2020,lanoy_2020,delory_2022}. Recently, the evolution of the dispersion curves in a stretched free strip~\cite{delory_2023b} has been studied and accurate predictions have been made. In this context, it is straightforward to compare elastography results with numerical predictions.

Here, using a simple silicon strip immersed in water and a simple ultrasound sequence, we suggest solutions to overcome the problems posed by viscoelasticity, waveguiding geometry and prestress. Anisotropy is naturally taken into account since the application of a prestress leads to extrinsic anisotropy for shear waves propagation in soft media~\cite{delory_2023a}.
By applying large deformations to the viscoelastic strip in different orientations, a wide range of phase velocities is measured. First, we identify the nature of the shear waves generated. Then, combining our previous work with the method described in~\cite{kiefer_2019}, we predict their dispersion curves. Finally, we are also able to capture the change of velocity with prestress. Our work tackles the above limitations, and paves the way for quantitative elastography.

\section{Experiment evidencing the limits of current shear wave elastography}
To begin, we want to explore the current limitations of elastography and see how they arise during an experiment. A simple silicone strip is placed in a water tank and acts as a waveguide for elastic waves.
Elastography experiments are performed using an Aixplorer\textsuperscript{TM} Multiwave ultrasound system and an XC6-1 curved array transducer from Supersonic Imaging. 
A standard Supersonic Shear wave Imaging (SSI) ultrasound sequence is used, consisting of 5 push lines, each composed of 4 push depths. After each push line, the transducer switches to the imaging mode (framerate of 1750 frames per second) to follow in real time the generated shear waves~\cite{bercoff_2004,deffieux_2008}.
\begin{figure*}
    \centering
    \includegraphics[width=15cm]{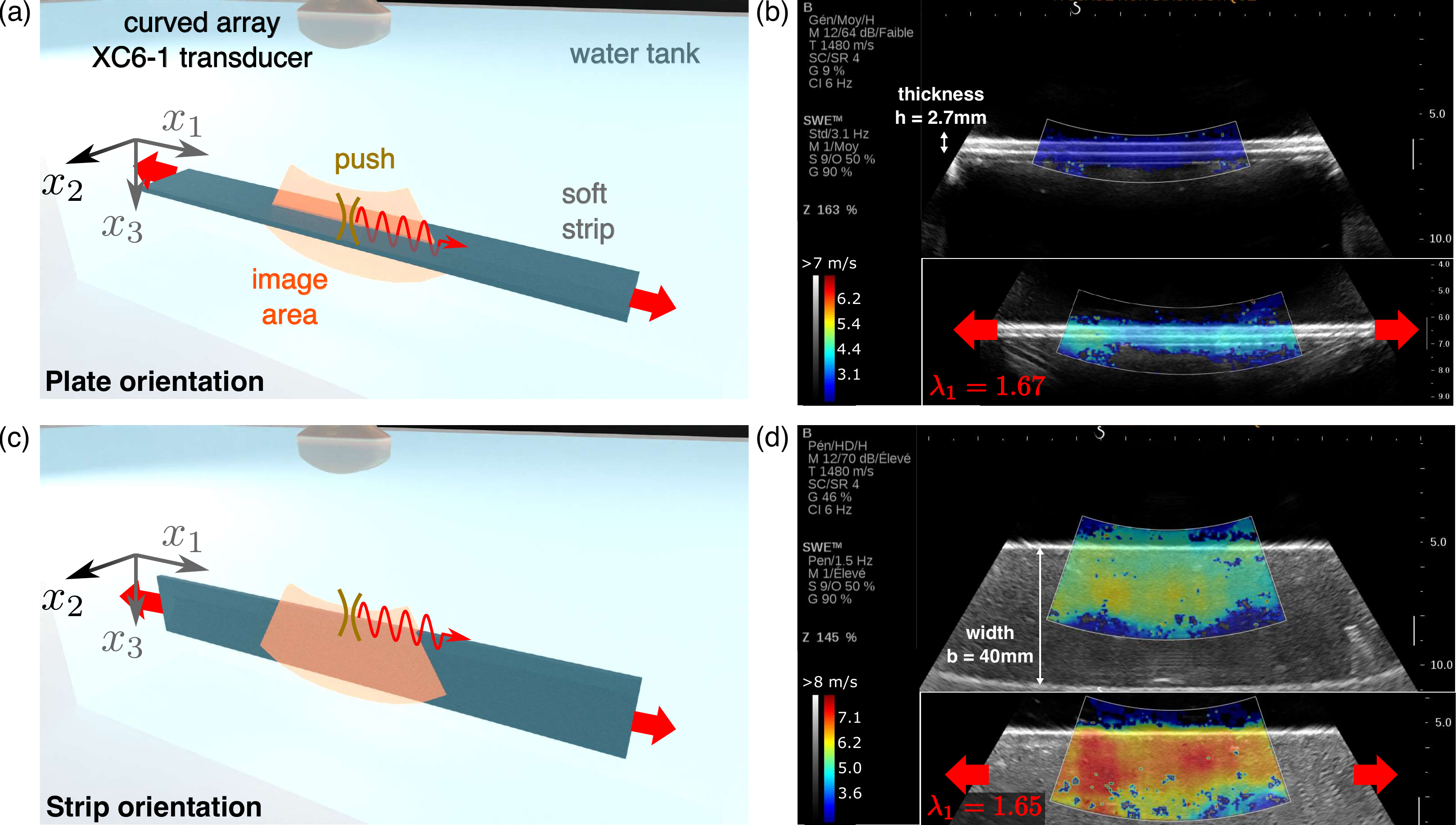}
    \caption{\textbf{Elastic waves are guided in two different orientations ---} Using the Supersonic Shear wave Imaging technique, shear waves are generated in a strip made of Ecoflex-0020 in two different orientations. In the plate orientation (respectively strip), the imaging plane $(e_1,e_3)$ cuts the strip along its thickness (a) (resp. along its width (c)). Typical results from the Aixplorer\textsuperscript{TM} ultrasound system are shown in (b) (resp. (d)) for an undeformed strip. The B-mode image is rendered in grayscale, while the velocity map is rendered in color. Then, the strip is stretched by a factor $\lambda_1$ in its length ($e_1$ direction) and the results are shown in the corresponding inset.}
    \label{fig:config}
\end{figure*}

The strip-shaped sample is made of an elastomer and prepared with thickness $h=2.7$~mm, width $b=4$~cm and length $L_0=20$~cm.
The elastomer is silicone Ecoflex-0020, a material commonly used in ultrasound imaging studies.
Today, the properties of Ecoflex are well known, both statically thanks to tensile tests~\cite{liao_2021}, and dynamically thanks to rheological measurements~\cite{liu_2014,kearney_2015,delory_2022}. It is soft enough to mimic biological tissues (YM of $\sim\!50$~kPa) and is preferred to agar gels for practical reasons (\textit{e.g.} it doesn't age).

The strip can be placed and examined in two different orientations as described in \fref{fig:config}. In both cases, the transducer array is parallel to the strip axis~$e_1$ and each push line generates a displacement along $e_3$.
In the plate (respectively strip) orientation depicted in \fref{fig:config}(a) (resp. \fref{fig:config}(c)), the imaging plane cuts the strip along its thickness (resp. width), as shown in the grayscale B-mode image in \fref{fig:config}(b) (resp. \fref{fig:config}(d)) and an out-of-plane (resp. in-plane) displacement is generated.

For the two orientations, the scanner provides the measured velocity of the shear waves within the strip. They are color-coded in \fref{fig:config}(b) and (d). Interestingly, the measured velocities are different: in the plate orientation, a velocity of $\sim\!3$~m/s is measured, while in the strip orientation, it is $\lesssim\!5$~m/s. Also, neither of these two values corresponds to the bulk shear velocity of the same elastomer, which would be around 5.3~m/s. As mentioned in the introduction, this is due to the fact that the strip acts as a waveguide and not as a bulk material. 
This is a first illustration that velocity is not a sufficient parameter to retrieve the stiffness of the material under consideration. While we can retrieve $E=3\rho {V_T}^2$ for a bulk wave, this is no longer possible for waves in the strip.

Now, the exact same sample is subjected to a uniaxial static stress along its length ($e_1$ direction) inducing an elongation of $\sim\!65\%$. The shear wave measurements are repeated for both orientations. The measured velocities are shown as insets in \fref{fig:config}(b) and (d). Both of them have increased to reach $\sim\!4$~m/s for the plate orientation and $\sim\!6.5$~m/s for the strip one.  
As introduced earlier, such changes in the velocity are due to the acoustoelastic effect~\cite{catheline_2003,gennisson_2007,destrade_2007,crutison_2022}. This is a second straightforward illustration of the limitations of quantitative elastography.

To summarize this part, we have taken a piece of soft material in the form of a thin strip. Shear wave elastography measurements with a commercial scanner yielded 4 different shear velocities for the same sample under different experimental conditions (orientation and initial stress). This is a significant challenge for a medical application designed to provide accurate quantitative stiffness measurements.
From a physical point of view, waveguiding and acoustoelasticity are the key phenomena to explain these variations. The goal of the next sections is to extract the material parameters from these measurements. 

\section{Waveguiding and dispersion}
In this section, we first take a closer look at the measured displacements to extract the dispersion curves of the waves. This is the so-called shear-wave spectroscopy technique which allows to capture the frequency-dependence of wave velocities~\cite{gennisson_2010,nguyen_2011,deffieux_2015}. Then, by solving for guided elastic waves in a strip, we can identify these waves, understand how they were generated and plot their dispersion curves.

\subsection{Experiment}
We use the Aixplorer\textsuperscript{TM} in research mode, which allows us to extract the full beamformed sequence of images after releasing a push line. The displacement field $u_3\left(x_1,x_3,t\right)$ is obtained by taking the phase of the correlation between successive images~\cite{bercoff_2004}.
Each image sequence is acquired five times and the resulting displacement fields are averaged in order to improve the signal-to-noise ratio.
\begin{figure*}
    \centering
    \includegraphics[width=15cm]{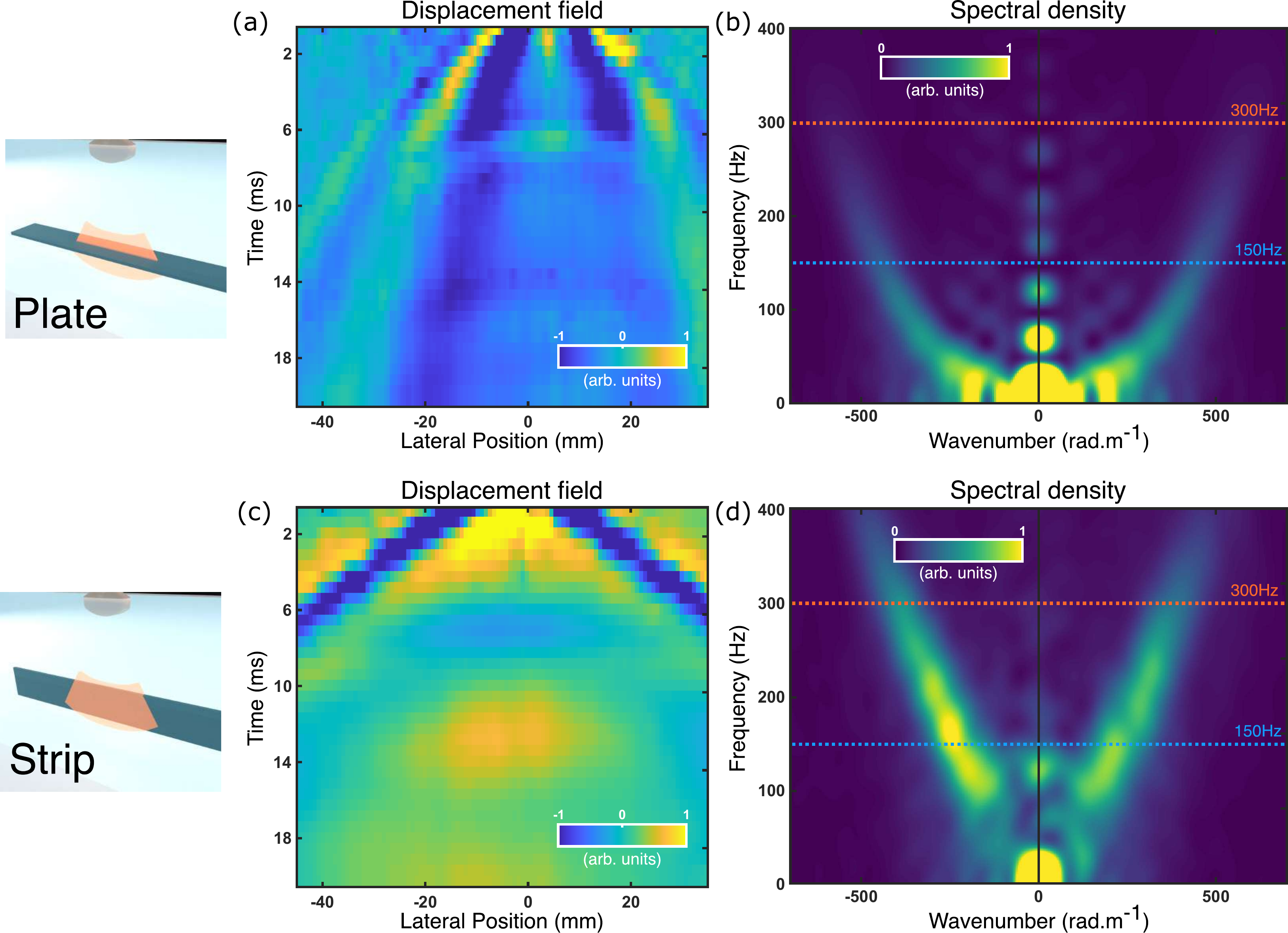}
    \caption{\textbf{Spectroscopy shear-wave elastography in both orientations ---} (a,c) For a given push line, a displacement field is obtained and averaged over 4 or 5 consecutive acquisitions. (b,d) The spatio-temporal Fourier transform gives the dispersion curves of the guided elastic waves in the given orientation. It is summed over 5 different push locations.}
    \label{fig:method}
\end{figure*}
The datasets consisting of 2D movies need to be reduced for visualization. For the plate orientation, the displacement is homogeneous along the thickness (direction $e_3$). Therefore, we decide to average the value along the thickness and obtain an averaged spatio-temporal displacement for each push line. The result for the push in the center of the scanned area is shown in \fref{fig:method}(a). For the strip orientation, the post-processing is slightly different since the displacement is no longer homogeneous along $e_3$.
In fact, only the displacement of the upper strip edge can be studied in an unbiased manner\footnote{%
The reason is that the longitudinal velocity used for imaging was assumed to be the same as that of water, \textit{i.e.} 1480~m/s, but sound propagates at about 1000~m/s in our material. This leads to two additional difficulties. First, the $x_3=ct$ axis is properly computed between the transducer and the first strip edge at $x_3=-b/2$, but in the strip, this axis is incorrectly estimated and the beamforming procedure is biased. This can be seen in \fref{fig:config}(d) where the second edge (at $x_3=+b/2$) appears curved at the bottom of the B-mode image. Second, the push focusing must also be degraded as we go deeper in the strip.}. %
The spatio-temporal displacement map corresponding to this top edge displacement is shown in \fref{fig:method}(c).

Both maps depict a localized displacement at the central position at time $t=0$ (top line) which then travels symmetrically towards the left and right directions with increasing time $t$. Comparing the two spatio-temporal displacement maps, we see that the two orientations give different results. And, confirming the previously measured velocity, the shear wave reaches the edges of the scanned area earlier in the strip orientation than in the plate orientation.
Moreover, in the plate orientation, short wavelengths seem to travel faster than long ones that have not reached the left edge at the late times of the presented image, indicating a dispersive behavior.
A last observation from these maps is that echoes appear on the abscissa of the push at 8~ms and 13~ms for the plate and strip orientations, respectively.

In order to evidence the waveguiding phenomenon within the strip at the origin of these observations, we propose to adopt the usual formalism of the community and extract their dispersion curves~\cite{royer2022elastic,auld1973acoustic}. By applying a spatial and temporal Fourier transform to the displacement map, a frequency versus wavenumber map of the same data can be obtained.
The magnitudes of such representations are normalized for each of the five push lines and averaged in \fref{fig:method}(b) and (c). Therefore, the results shown are not local but correspond to an average over the entire scanned area. 

Again, the two maps exhibit different behavior. First, while the strip orientation gives a linear dispersion curve, the plate orientation gives a convex one. This is another way of evidencing the dispersive nature of the propagation in the plate orientation.
Second, the intensities are not evenly distributed with the frequency in the two maps. In the plate orientation, the intensities are high in the low-frequency range and decreases rapidly with increasing frequency, almost disappearing around 300~Hz. On the contrary, energy is found at higher frequencies in the strip orientation, with a maximum intensity around 150~Hz (leaving aside the zero-frequency spot).
Finally, we notice some spots for certain frequencies on the $k=0$ axis, in both orientations. These spots are due to the above mentioned echoes from the lateral edges of the strip.
In fact, in a waveguide, these echoes usually materialize as cut-off frequencies in the dispersion diagram (see Appendix~A for more details) and are characterized by $k=0$.

\subsection{Theory}
For a more thorough understanding of the propagation, we begin the theoretical interpretation by neglecting viscoelastic effects and nonlinear mechanics.
To this end, we use finite element simulations (here with COMSOL Multiphysics) to search for the dispersion curves of guided elastic waves in a strip immersed in water. A strip of thickness $h=2.7$~mm and width $b=4$~cm is considered with a density of 1.07~g/cm$^3$, a longitudinal velocity of 1000~m/s \cite{delory_2022} and a transverse velocity of 5.31~m/s. In this coupled simulation, the motion of the strip induces displacements of the surrounding water. Perfectly Matched Layers (PML) were used to mimic a non-reflecting infinite water domain. The corresponding dispersion curves for both orientations are shown in \fref{fig:modeShape} (see Appendix~A for diagrams containing all modes).
In the plate orientation, the dispersion curve corresponding to the experimental one is displayed as a thick red line. Its power-law behavior is a known characteristic of a bending mode~\cite{delory_2023b}, and indeed, the mode displacements shown as insets for two different frequencies reveal a bending motion of the strip. This is consistent with the SSI scenario envisioned previously, where the strip was pushed down in the plate orientation.
Note that the effect of the strip width begins to be important at higher frequencies, as the displacement profile becomes inhomogeneous in the width direction at 100~Hz.

Alternatively, the pushes in the strip orientation generate in-plane displacements in the strip. Many modes can propagate in such a strip and comparing the theoretical dispersion curves with the experimental one, as well as their displacement profiles, we conclude that the generated mode is the first antisymmetric mode\footnote{Or a combination of the first antisymmetric mode and the first symmetric mode at higher frequencies, which we denote as "edge wave".}. The dispersion curve of this mode is plotted as a thick blue line in \fref{fig:modeShape} with negative wavenumbers. 
Again, the displacement profile of this mode at 10~Hz and 300~Hz is compatible with an excitation at the top edge of the strip. At 10~Hz, this mode is actually dispersive since it is really resembles a bending of the strip, but this time in its width. As the frequency increases, the wavelength decreases and the guide width becomes larger than the shear wavelength: most of the mode energy is now confined to the edge. %
Remembering that the experimental spectral intensities are high for frequencies above 100~Hz (\fref{fig:method}), a wave propagating along the strip edge is expected with fast amplitude decay along $e_3$~\cite{delory_2022}.
\begin{figure*}
    \centering
    \includegraphics[width=13cm]{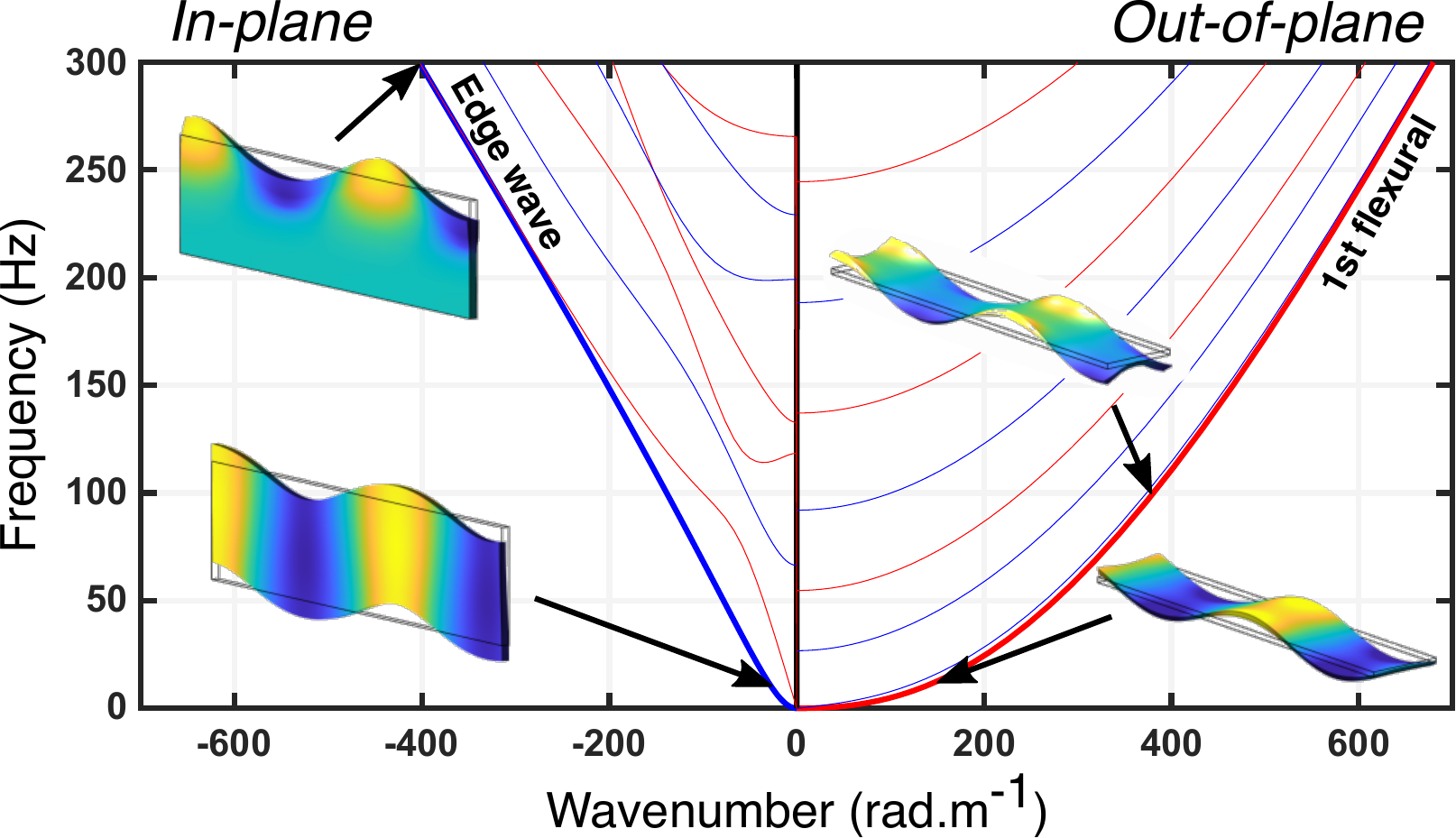}
    \caption{\textbf{Simplified dispersion diagram for a strip in water ---} A linear elastic strip is considered with thickness $h=2.7$~mm, width $b=4$~cm, transverse velocity $V_T=5.31$~m/s and longitudinal velocity $V_L=1000$~m/s. Two sets of dispersion curves are plotted, the out-of-plane (or bending) modes for positive wavenumbers, and the in-plane modes for negative wavenumbers.
    Other dispersion curves are also plotted as thin lines. The displacement components $u_3$ are displayed for both modes at 10~Hz. The bending mode is also shown at 100~Hz where the effect of the lateral boundaries begins to show.
    The edge wave is displayed at 300~Hz and is the sum of the first symmetric and antisymmetric in-plane modes. The strip axis is not to scale.}
    \label{fig:modeShape}
\end{figure*}

Regarding the echoes observed in \fref{fig:method}(a) and (c) and the corresponding spots on the $k=0$ axis in \fref{fig:method}(b) and (d), such a simulation allows to attribute them to the cut-off frequencies of higher order modes.
Note that the first cut-off frequency in the in-plane mode dispersion diagram does not appear experimentally in \fref{fig:method}(d). It would correspond to a shear wave propagating back-and-forth in the $e_3$ direction. However, since the generated displacement is polarized in the $e_3$ direction, no shear wave propagates in the $e_3$ direction, in the experiment.
In contrast, the second cut-off frequency corresponds to a wave of longitudinal appearance and is generated precisely because it corresponds to a displacement and propagation in the $e_3$ direction.

Next, we consider the viscoelasticity of the material. This affects the dispersion curves shown in \fref{fig:modeShape}. Theoretically, the edge mode is almost non-dispersive. However, by extracting the experimental phase velocity $V_\phi$ of this mode, we obtain 4~m/s at 150~Hz and 4.8~m/s at 300~Hz.
This dispersion is due to the viscoelasticity of the medium, \textit{i.e.} its frequency-dependent material properties. From a mathematical point of view, viscoelasticity includes the rheological model of the material. Usually, the latter is determined by using a dedicated device such as a plate-plate rheometer. For our material, the best model is a fractional Kelvin-Voigt model where the complex shear modulus writes: $$\mu\left(\omega\right) = \mu_0 \left[1+\left(\rmi \omega \tau \right)^n \right].$$
This viscoelastic model is becoming widespread in the soft mechanics literature~\cite{sharma_2023} and is recommended to model the behavior of soft tissues~\cite{parker_2019} as well as the Ecoflex silicone considered here~\cite{yasar_2013,liu_2014,delory_2022}.

A difficulty then arises, since we do not have a numerical tool to take into account the fractional viscoelastic model in a strip immersed in water. To overcome this, we distinguish between the two orientations and solve two different problems. For the plate orientation, the dispersion of the first bending mode is almost identical to that in a plate (see Appendix~A for details). Therefore, we can solve the simpler problem of a plate immersed in water to calculate the dispersion. Next, a method is used that allows the implementation of frequency-dependent parameters~\cite{kiefer_2019}.
For the strip orientation, the problem is actually not easy to solve because there is no similar method for a strip immersed in water. However, the same procedure exists for a free strip~\cite{delory_2023b}, \textit{i.e.} not immersed in water, and again allows to use frequency-dependent parameters. In such a scenario, where the presence of water is neglected, the dispersion curves of the in-plane guided modes are slightly modified.
To quantify the error, we performed COMSOL simulations comparing the dispersion curves with/without water in the dispersive case (see Appendix~A). The velocity is reduced by a factor of 1.15 when the coupling of water is added. Subsequently, we will increase the wavenumbers by a factor of 1.15 to simulate the influence of water when calculating solutions in its absence.
Note that a close factor appears when modeling surface waves at the interface between an incompressible elastic medium and air or water, as indicated in~\cite{kirby_2017,pelivanov_2019}: $$V_{\rm{air/solid~ interface}}/V_{\rm{water/solid~interface}} = V_{\rm{Rayleigh}}/V_{\rm{Scholte}} = 1.13.$$

Finally, these two methods are used to make predictions for the phase velocities of the two modes of interest. In this way, viscoelasticity no longer becomes a problem, but rather an asset that we can use to solve an inverse problem and determine the complex shear modulus $\mu\left(\omega\right)$, \textit{i.e.} the rheology of the material.
Note that this is \textit{a priori} not straightforward since the dispersion also originates from the waveguiding, especially for the bending mode in the plate orientation, but this can still be overcome using the previously described methods.

In summary, this theoretical part has allowed to evidence the two guided modes that are excited in the plate and strip orientations. Their dispersion relations can be predicted for all frequencies.
Overall, the results presented in this section exhibit both the effect of the guiding geometry and the frequency dependence of the parameters. The measured phase velocity $V_\phi$ at 150~Hz or 300~Hz would lead to different values for the YM $E=3\rho {V_\phi}^2$, as indicated in \fref{fig:results}(c). Thanks to our theoretical modeling, we can actually predict the correct dispersion and achieve better material characterization than when considering a bulk propagation.

\section{Stretching and acoustoelastic effect}
Knowing the nature of the generated waves and their dispersion curves, we now focus on the effect of stretching the strip. We believe that this effect can correspond to many physiological situations: the muscles and tendons stretch themselves, the arteries support pulsatile waves \cite{baranger_2023}, or more generally the ultrasound sonographer can compress the tissues under investigation.
Here, we aim at showing that one can predict the shear wave propagation by knowing the material parameters, or vice versa, measuring the same tissue under different loads allows to obtain more data and better characterize the material.

\subsection{Experiment}
The idea is to deform the sample before measuring the propagation of shear guided waves. Experimentally, we impose a new length to the strip by stretching its two extremities in the $e_1$ direction. The deformation is characterized in terms of stretch ratios along all directions $\left(\lambda_1, \lambda_2, \lambda_3\right)$. Assuming an incompressible material and a uniaxial elongation along $e_1$, we have $\lambda_2 = \lambda_3 = 1/\sqrt{\lambda_1}$.
The ultrasound scanner is used for tracking the shear wave propagation, and the experiment is repeated for several values of $\lambda_1$ ranging from 1 (undeformed) to 1.67 (insets of \fref{fig:config}(b) and (d)). 
For each stretch ratio and both orientations, the spatio-temporal displacement maps are extracted by correlation between two consecutive images. The spatio-temporal Fourier transform is applied in order to obtain the frequency versus wavenumber representation of the same data\footnote{Note that again the maps are averaged over 5 distinct push positions.}. Finally, the maximum for each frequency is detected in order to draw the extracted dispersion relation.
\begin{figure*}
    \centering
    \includegraphics[width=\textwidth]{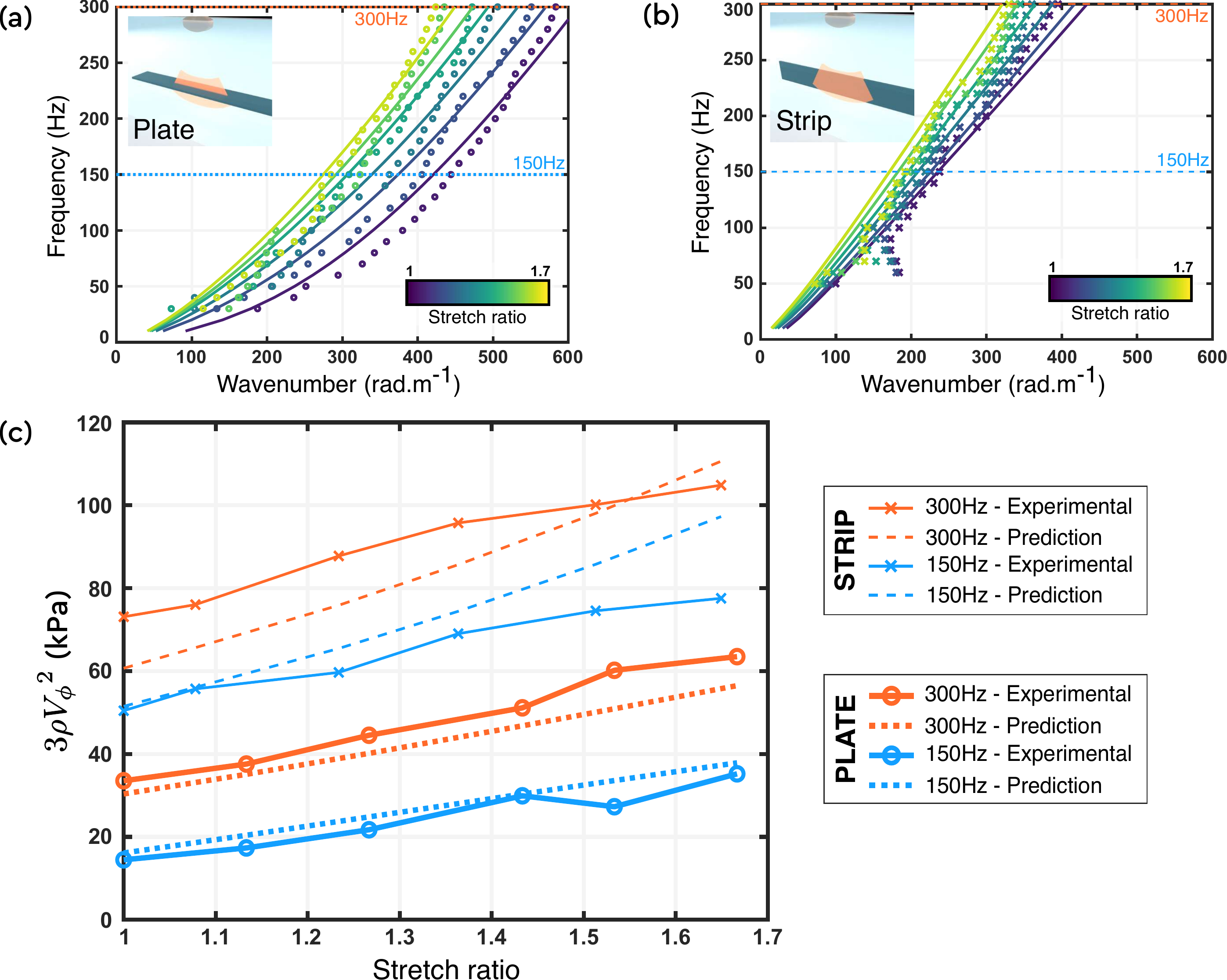}
    \caption{\textbf{Measured dispersion curves for various stretch ratios ---} For each stretch ratio, wavenumbers are extracted from the dispersion curve at certain frequencies ranging from 30 to 300~Hz and plotted as circles for the plate orientation (a) and crosses for the strip orientation (b). Predictions are added as solid lines and match both the dependence in frequency and stretching. (c) In addition, by looking for an appropriate Young modulus $E$, one should compute the quantity $3\rho {V_{\phi}}^2$. The multiple values, obtained with a single sample, range from 15 to 105~kPa.}
    \label{fig:results}
\end{figure*}

The results of the treated experiments are summarized as symbols in \fref{fig:results}(a) (resp. (b)) for the plate (resp. strip) orientation. The color coding stands for the stretch ratio $\lambda_1$. The dark blue points correspond to the same data as in \fref{fig:method} in the previous section. Frequencies above 300~Hz are discarded in the plate orientation since there is almost no signal above, and the same for frequencies below 50~Hz in the strip orientation. 

In both scenarios, the application of a static stress tends to increase the slopes of the dispersion curves: the greater the stretching, the higher the frequency for a given wavenumber. In other words, the velocity of the wave propagating along the stretched direction is increased. This is a relatively intuitive behavior that can be easily experienced with a stretched rope. However, while the general trend is easy enough to conceptualize, the details of the increase are harder to grasp.

To better assess this evolution, the phase velocities $V_\phi$ are extracted at 150~Hz and 300~Hz and the quantity $3\rho {V_{\phi}}^2$, similar to a Young modulus, is plotted (symbols) as a function of the stretch ratio in \fref{fig:results}(c). This plot now evidences a linear increase in the apparent YM with the stretch ratio, and the increases are similar for both orientations. However, the main conclusion to be drawn is that a wide range of velocities is obtained for the same experimental sample. The standard technique that consists in defining the YM directly from the measured velocity would lead to values varying from 15~kPa to 105~kPa.
One idea would be to take advantage of all these measurements to better characterize the medium under investigation and to build an inverse problem to infer the material constants.

\subsection{Theory}
The experimentally evidenced effect of stretching is modeled in the following.
As mentioned earlier, the changes in phase velocities with stretching are commonly referred to as the acoustoelastic effect and are particularly important in soft media that are highly-deformable. This theory relies on first modelling a large static deformation and then considering the propagation of small perturbations. This has already been investigated in a similar elastomer with guided waves in a plate~\cite{delory_2023a} and in a strip~\cite{delory_2023b}. It has been shown that both hyperelastic and viscoelastic material properties must be taken into account. 

At high deformations, it is essential to consider the deviation from Hooke's law. Popular models to account for this are hyperelastic laws, which are based on a strain energy density function $W$. Among the many possible empirical laws, the simplest (and oldest) one is known as the Mooney-Rivlin model~\cite{rivlin1948large}. Although simple, it has allowed correct predictions for Ecoflex in~\cite{delory_2023a} for elongation ratio $\lambda_1$ lower than $2$ as considered here. Alternatively, one could choose any of the other hyperelastic models used to model biological tissues~\cite{wex_2015,chagnon_2015}.

Once the mechanical properties of the highly deformed material are understood, the wave perturbation in this new medium is considered. The latter is time-dependent and viscoelasticity must be accounted for. 
Overall, the acoustoelastic theory can incorporate both the hyperelastic and the viscoelastic  models~\cite{delory_2023a}. The Cauchy stress tensor is replaced by the sum of the usual hyperelastic contribution and an additional time-dependent one. Finally, a modified wave equation is obtained, provided that we use an equivalent elasticity tensor $\mathbf{C}^{\omega}$.
See Appendix~B for more details on the calculations and this equivalent elasticity tensor. Note that when dealing with anisotropic material, additional invariants should be considered to write the strain energy density function~\cite{balzani_2006,peyraut_2010,li_cao_2017,mukherjee_2022}.

The last step consists in applying the boundary conditions to determine the dispersion relations of the guided modes. The procedure is the same as before except that we adopt the modified tensor $\mathbf{C}^{\omega}$ and the deformed geometry. The coupling with the surrounding water must also be taken into account. This step is probably the most complicated one, although it is fully solved for plates~\cite{li_cao_2017,kiefer_2019} and we believe it goes beyond the scope of this article. For the theoretical predictions presented in \fref{fig:results}, we proceed similarly as before, \textit{i.e.} solving the problem of a plate immersed in water to obtain the dispersion curve of the first bending wave in a strip; and increasing the wavenumbers by a factor of 1.15 to get the dispersion curve of the edge wave.
For all predictions, we considered a strip of Ecoflex-0020, with thickness $h=2.7$~mm and width $b=4$~cm. We use the rheological parameters $\mu_0=15$~kPa, $\tau=1000$~µs and $n=0.33$; and the hyperelastic parameters $\alpha=0.29$ and $\beta'=0.29$ (see Appendix~B). These parameters were chosen after manually adjusting the dispersion curves by fixing the values for $\alpha$ and $\beta'$ since they had already been assessed for a plate made of a similar material~\cite{delory_2023a} and also gave good predictions in a strip~\cite{delory_2023b}.

\subsection{Discussion}
For both orientations, the predictions provided by our approach are very satisfactory in \fref{fig:results}(a) and (b).
Similarly, the increases in velocity with elongation are also well understood, whether in the plate or strip orientation, at an intermediate frequency of 150~Hz or at a higher frequency of 300~Hz, as seen in \fref{fig:results}(c). Moreover, the slope of these increases is well predicted.

Some differences still persist in \fref{fig:results}. This is because the exact problem is not solved in either orientation, errors in \fref{fig:results}(b) are probably due to the oversimplified modeling using a free strip in air with a correction factor.
Secondly, errors can also be attributed to the non-linear material model since it has only been validated for Ecoflex-0030 and not for Ecoflex-0020~\cite{delory_2023a}, and also because the hyperelastic model used (Mooney-Rivlin) remains a weakly non-linear elastic model.

In the end, we see that with a single strip, classical elastography yields Young moduli ranging from 15~kPa to 105~kPa. This is a very wide range of values, and these experiments clearly highlight the limitations of shear wave elastography. More importantly, we are able to fully explain this wide range of values.

Based on the presented measurements, we can imagine the implementation of an inverse method to probe the rheological and hyperelastic parameters of the material under study.
And to go even further, ultrasound images can be used to monitor the evolution of the geometric parameters ($h,b$) with the applied prestress. These measurements should be carried out in conjunction with the evolution of the cut-off frequencies, which provide almost direct information on these geometric parameters.

\section{Conclusion}
The influence of frequency, geometry and static deformation in elastography is captured in this work, using a single material and a simple experimental method.
We show that neglecting these effects can lead to a wide range of incorrect Young moduli and we provide solutions for modeling and understanding the guided waves generated in supersonic shear wave imaging.
The procedure used in this work can be adapted to other material models, including anisotropic ones. Furthermore, a generalization to other guiding geometries could be performed.
Finally, the inverse problem could be solved to infer the hyperelastic and viscoelastic properties from the measured dispersion curves.

\ack{}
We would like to acknowledge the valuable help of Arthur Le Ber and Flavien Bureau for the use of the Aixplorer\textsuperscript{TM} ultrasound system.
This work has received support under the program “Investissements d’Avenir” launched by the French Government and partially supported by the Simons Foundation/Collaboration on Symmetry-Driven Extreme Wave Phenomena. A.D. acknowledges funding from French Direction Générale de l’Armement.

\newpage
\section*{Appendix A: Full dispersion curves of guided elastic waves in a strip with free boundary conditions}
\begin{figure*}[b]
    \centering
    \includegraphics[width=13cm]{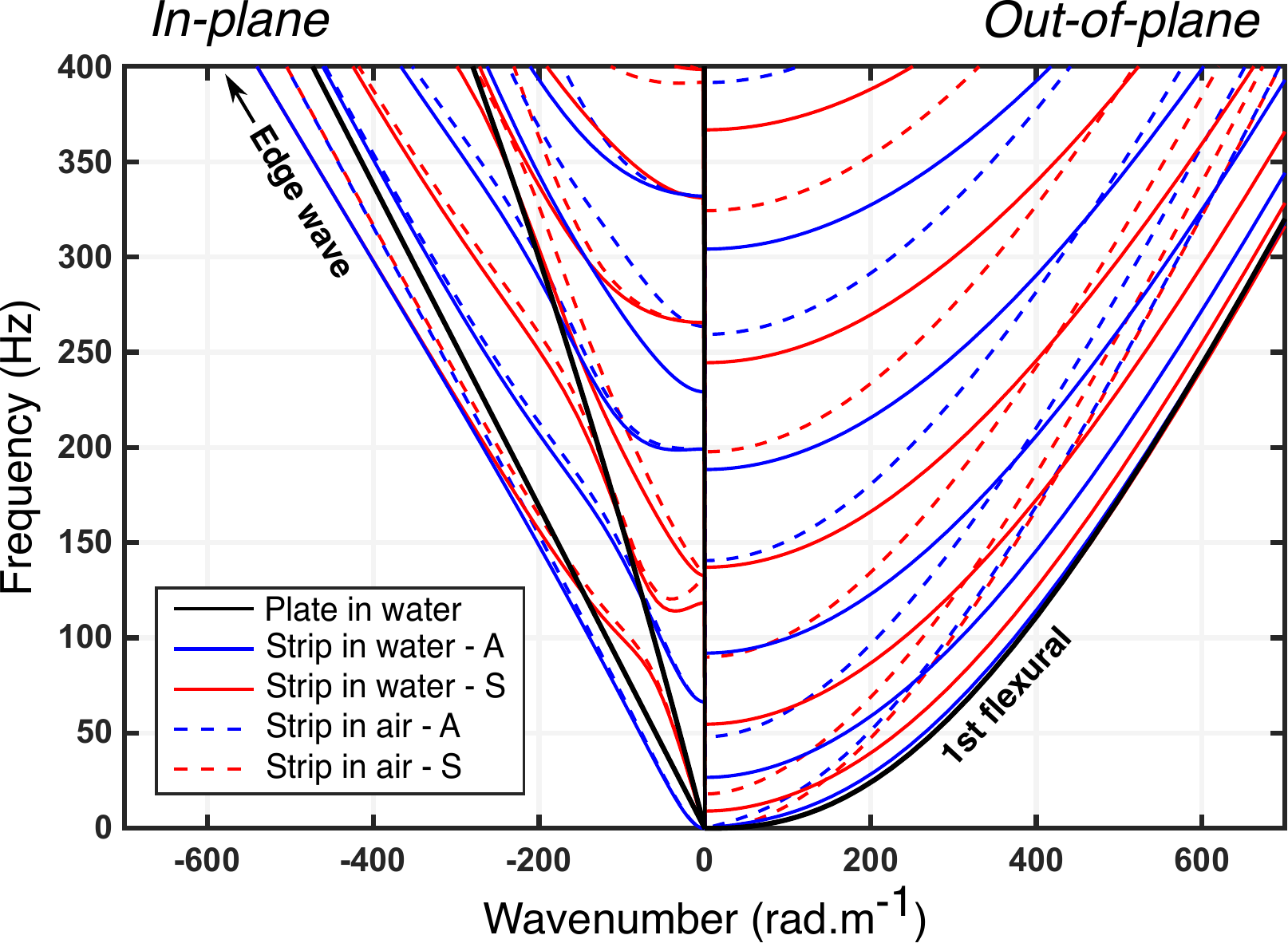}
    \caption{\textbf{Dispersion diagrams for a strip in air and water ---} A purely elastic strip is considered with thickness $h=2.7$~mm, width $b=4$~cm and transverse velocity $V_T=5.31$~m/s. The dispersion curves are split in two parts: the out-of-plane (or bending) modes mainly polarized along the thickness axis, and the in-plane modes mainly polarized in the plane of the width and propagation axis. Two dispersion diagrams are superimposed: with (full lines) and without (dashed lines) coupling with water. When solving for those elastic guided waves, it is common to distinguish between antisymmetric (A in blue) or symmetric (S in red) with respect to the strip axis.}
    \label{fig:appA}
\end{figure*}

The elastography technique allows us to study two guided elastic modes in a soft strip immersed in water, but it is important to remember that there are many more modes that can propagate in such a strip. Here, we solve for the full dispersion diagram of a soft strip using eigenfrequency analysis from COMSOL finite element software (\fref{fig:appA}). The strip has the same dimensions as before, thickness $h=2.7$~mm and width $b=4$~cm.
However, as mentioned in the main text, it is not easy to take into account both the 3D geometry, the viscoelasticity and the prestress. In particular, when considering the viscoelasticity and the prestress using an equivalent elasticity tensor as described in Appendix~B, it becomes difficult to implement adequate boundary conditions, neither to solve for the wavenumber $k$ nor for the frequency. This leads us to perform simulations using a simple homogeneous, isotropic and purely elastic material with a transverse velocity $V_T=5.31$~m/s.

In \fref{fig:appA}, the dispersion diagrams of a soft strip in air or immersed in water are compared. The coupling with water has a significant effect on the bending modes which displace an additional amount of water, compared to the strip in air, adding inertia and lowering the dispersion curves. On the other hand, the effect on the in-plane guided modes is less important but still significant. In particular, the edge wave velocity is reduced by a factor of 1.15 for such a material.

Finally, the dispersion curves of elastic guided waves in a plate are added. Looking at the first bending mode that can propagate in a plate (black solid line on the right part of \fref{fig:appA}), one notices that it almost coincides with the first bending mode that can propagate in a strip\footnote{Note that this would not be true when working at lower frequencies (\textit{i.e.} great wavelengths).}. This remark is important because it allows us to analyze only the simple plate problem when studying the generated wave in the plate orientation.

\section*{Appendix B: Complete visco-hyperelastic constitutive law of Ecoflex}
To describe waves in a pre-stressed viscoelastic body, an incremental approach~\cite{delory_2023a} is built as described by Ogden and Destrade~\cite{ogden_1997,destrade_2007}. The main result of this theory is that all the non-linearities can be included in a new tensor $\ten{C^\omega}$ and a wave equation is still obtained:
\begin{equation}
    C^\omega_{jikl} \frac{\partial^2 u'_k}{\partial x_j\partial x_l} + \rho \omega^2 u'_i = 0
\end{equation}
with $\mathbf{u'}(\mathbf{x},\omega)=\mathbf{x'}-\mathbf{x}$ an incremental monochromatic displacement and $\ten{C^\omega}$ the modified elasticity tensor that depends on the strain energy density function $W$, the viscoelastic model and the stretch ratios~$\lambda_i$. $\ten{C^\omega}$ actually decomposes as:
\begin{equation}
    C^\omega_{ijkl} = C^0_{ijkl} + \mu_0\left(\rmi\omega\tau\right)^n \left(1 +\beta'\,\frac{\lambda_i^2+\lambda_j^2-2}{2}\right) \left(\delta_{ik}\delta_{jl}+\delta_{il}\delta_{jk}\right)
\end{equation}
where $C^0_{ijkl}$ is the equivalent stiffness tensor for the usual acoustoelastic effect (without viscoelasticity) of an hyperelastic solid. Here a compressible Mooney-Rivlin solid is selected, with $W$ that writes:
\begin{equation}
    W = \frac{\mu_0}{2}\Bigg[(1-\alpha)\left(\frac{I_1}{J^{2/3}}-3\right)+\alpha\left(\frac{I_2}{J^{4/3}}-3\right)\Bigg] + \frac{K}{2}(J-1)^2
\end{equation}
with $I_1 = \lambda_1^2 + \lambda_2^2 + \lambda_3^2$, $I_2 = \lambda_2^2\lambda_3^2 + \lambda_1^2\lambda_3^2 + \lambda_1^2\lambda_2^2$ and $J = \lambda_1\lambda_2\lambda_3$. Given this equation for $W$, one can write the coefficients of the tensor $\ten{C^0}$:
\begin{eqnarray}
    C^0_{iijj} &= \frac{\lambda_i\lambda_j}{J} \, W_{ij}&\quad \\
    C^0_{ijji} &= \frac{\lambda_i^2}{J} \, \frac{\lambda_i W_i - \lambda_j W_j}{\lambda_i^2-\lambda_j^2} &\quad(i\neq j, \lambda_i\neq\lambda_j)\\
    C^0_{ijji} &= \frac{C^0_{iiii}-C^0_{iijj}+\lambda_i W_i/J}{2} &\quad(i\neq j, \lambda_i=\lambda_j)\\
    C^0_{ijij} &= \frac{\lambda_i\lambda_j}{J} \, \frac{\lambda_j W_i - \lambda_i W_j}{\lambda_i^2-\lambda_j^2} &\quad(i\neq j, \lambda_i\neq\lambda_j)\\
    C^0_{ijij} &= \frac{C^0_{iiii}-C^0_{iijj}-\lambda_i W_i/J}{2} &\quad(i\neq j, \lambda_i=\lambda_j)
\end{eqnarray}
where $\displaystyle W_i = \frac{\partial W}{\partial \lambda_i}$ and $\displaystyle W_{ij} = \frac{\partial^2 W}{\partial \lambda_i \partial \lambda_j}$.
Here, formulas are slightly different from the ones you can find in books~\cite{ogden_1997,destrade_2007} because the dot products convention are different. To go from their definition to the one presented in this work, there is a simple permutation to accomplish for the last 2 indices. At the end, the wave equation to be solved is the same.

\section*{References}

\begin{thebibliography}{10}

\bibitem{sandrin_2003}
Laurent Sandrin, Bertrand Fourquet, Jean-Michel Hasquenoph, Sylvain Yon,
  Céline Fournier, Frédéric Mal, Christos Christidis, Marianne Ziol, Bruno
  Poulet, Farad Kazemi, Michel Beaugrand, and Robert Palau.
\newblock Transient elastography: a new noninvasive method for assessment of
  hepatic fibrosis.
\newblock {\em Ultrasound in Medicine \& Biology}, 29(12):1705--1713, 2003.

\bibitem{asbach_2010}
Patrick Asbach, Dieter Klatt, Beate Schlosser, Michael Biermer, Marion Muche,
  Anja Rieger, Christoph Loddenkemper, Rajan Somasundaram, Thomas Berg, Bernd
  Hamm, Juergen Braun, and Ingolf Sack.
\newblock Viscoelasticity-based {Staging} of {Hepatic} {Fibrosis} with
  {Multifrequency} {MR} {Elastography}.
\newblock {\em Radiology}, 257(1):80--86, 2010.

\bibitem{deffieux_2015}
Thomas Deffieux, Jean-Luc Gennisson, Laurence Bousquet, Marion Corouge, Simona
  Cosconea, Dalila Amroun, Simona Tripon, Benoit Terris, Vincent Mallet,
  Philippe Sogni, Mickael Tanter, and Stanislas Pol.
\newblock Investigating liver stiffness and viscosity for fibrosis, steatosis
  and activity staging using shear wave elastography.
\newblock {\em Journal of Hepatology}, 62(2):317--324, February 2015.

\bibitem{kennedy_2018}
Paul Kennedy, Mathilde Wagner, Laurent Castéra, Cheng~William Hong, Curtis~L.
  Johnson, Claude~B. Sirlin, and Bachir Taouli.
\newblock Quantitative {Elastography} {Methods} in {Liver} {Disease}: {Current}
  {Evidence} and {Future} {Directions}.
\newblock {\em Radiology}, 286(3):738--763, March 2018.

\bibitem{sinkus_2005}
Ralph Sinkus, Mickael Tanter, Tanja Xydeas, Stefan Catheline, Jeremy Bercoff,
  and Mathias Fink.
\newblock Viscoelastic shear properties of in vivo breast lesions measured by
  {MR} elastography.
\newblock {\em Magnetic Resonance Imaging}, 23(2):159--165, 2005.

\bibitem{barr_2012}
Richard~G. Barr and Zheng Zhang.
\newblock Effects of {Precompression} on {Elasticity} {Imaging} of the
  {Breast}.
\newblock {\em Journal of Ultrasound in Medicine}, 31(6):895--902, 2012.

\bibitem{barr_2019}
Richard~Gary Barr.
\newblock Future of breast elastography.
\newblock {\em Ultrasonography}, 38(2):93--105, 2019.

\bibitem{correas_2013}
J.~M. Correas, A.~M. Tissier, A.~Khairoune, G.~Khoury, D.~Eiss, and
  O.~Hélénon.
\newblock Ultrasound elastography of the prostate: {State} of the art.
\newblock {\em Diagnostic and Interventional Imaging}, 94(5):551--560, 2013.

\bibitem{cantisani_2015}
Vito Cantisani, Hektor Grazhdani, Elena Drakonaki, Vito D’Andrea, Mattia
  Di~Segni, Erton Kaleshi, Fabrizio Calliada, Carlo Catalano, Adriano Redler,
  Luca Brunese, Francesco~Maria Drudi, Angela Fumarola, Giovanni Carbotta,
  Fabrizio Frattaroli, Nicola Di~Leo, Mauro Ciccariello, Marcello Caratozzolo,
  and Ferdinando D’Ambrosio.
\newblock Strain {US} {Elastography} for the {Characterization} of {Thyroid}
  {Nodules}: {Advantages} and {Limitation}.
\newblock {\em International Journal of Endocrinology}, 2015:e908575, 2015.

\bibitem{elgeti_2014}
Thomas Elgeti and Ingolf Sack.
\newblock Magnetic {Resonance} {Elastography} of the {Heart}.
\newblock {\em Current Cardiovascular Imaging Reports}, 7(2):9247, 2014.

\bibitem{sinkus_2014}
Ralph Sinkus.
\newblock Elasticity of the {Heart}, {Problems} and {Potentials}.
\newblock {\em Current Cardiovascular Imaging Reports}, 7(9):9288, 2014.

\bibitem{hansen_2015}
Hendrik~H.G. Hansen, Mathieu Pernot, Simon Chatelin, Mickael Tanter, and
  Chris~L. de~Korte.
\newblock Shear wave elastography for lipid content detection in transverse
  arterial cross-sections.
\newblock In {\em 2015 {IEEE} {International} {Ultrasonics} {Symposium}
  ({IUS})}, pages 1--4, 2015.

\bibitem{khan_2018}
Saad Khan, Faisal Fakhouri, Waqas Majeed, and Arunark Kolipaka.
\newblock Cardiovascular magnetic resonance elastography: {A} review.
\newblock {\em NMR in Biomedicine}, 31(10):e3853, 2018.

\bibitem{pruijssen_2020}
Judith~T. Pruijssen, Chris~L. de~Korte, Iona Voss, and Hendrik H.~G. Hansen.
\newblock Vascular {Shear} {Wave} {Elastography} in {Atherosclerotic}
  {Arteries}: {A} {Systematic} {Review}.
\newblock {\em Ultrasound in Medicine \& Biology}, 46(9):2145--2163, 2020.

\bibitem{prado_2018}
Rui Prado-Costa, João Rebelo, João Monteiro-Barroso, and Ana~Sofia Preto.
\newblock Ultrasound elastography: compression elastography and shear-wave
  elastography in the assessment of tendon injury.
\newblock {\em Insights into Imaging}, 9(5):791--814, 2018.

\bibitem{farron_2009}
Joe Farron, Tomy Varghese, and Darryl~G. Thelen.
\newblock Measurement of {Tendon} {Strain} {During} {Muscle} {Twitch}
  {Contractions} {Using} {Ultrasound} {Elastography}.
\newblock {\em IEEE transactions on ultrasonics, ferroelectrics, and frequency
  control}, 56(1):27--35, 2009.

\bibitem{mifsud_2023}
Tiziana Mifsud, Alfred Gatt, Kirill Micallef-Stafrace, Nachiappan Chockalingam,
  and Nat Padhiar.
\newblock Elastography in the assessment of the {Achilles} tendon: a systematic
  review of measurement properties.
\newblock {\em Journal of Foot and Ankle Research}, 16(1):23, 2023.

\bibitem{winn_2016}
Naomi Winn, Radhesh Lalam, and Victor Cassar-Pullicino.
\newblock Sonoelastography in the musculoskeletal system: {Current} role and
  future directions.
\newblock {\em World Journal of Radiology}, 8(11):868--879, 2016.

\bibitem{paluch_2016}
{\L}ukasz Paluch, Ewa Nawrocka-Laskus, Janusz Wieczorek, Bartosz Mruk,
  Ma{\l}gorzata Frel, and Jerzy Walecki.
\newblock Use of {Ultrasound} {Elastography} in the {Assessment} of the
  {Musculoskeletal} {System}.
\newblock {\em Polish Journal of Radiology}, 81:240--246, May 2016.

\bibitem{taljanovic_2017}
Mihra~S. Taljanovic, Lana~H. Gimber, Giles~W. Becker, L.~Daniel Latt, Andrea~S.
  Klauser, David~M. Melville, Liang Gao, and Russell~S. Witte.
\newblock Shear-{Wave} {Elastography}: {Basic} {Physics} and {Musculoskeletal}
  {Applications}.
\newblock {\em RadioGraphics}, 2017.

\bibitem{davis_2019}
Leah~C. Davis, Timothy~G. Baumer, Michael~J. Bey, and Marnix van Holsbeeck.
\newblock Clinical utilization of shear wave elastography in the
  musculoskeletal system.
\newblock {\em Ultrasonography}, 38(1):2--12, 2019.

\bibitem{ormachea_2020}
J.~Ormachea and K.~J. Parker.
\newblock Elastography imaging: the 30 year perspective.
\newblock {\em Physics in Medicine \& Biology}, 65(24):24TR06, 2020.

\bibitem{zviet_2022}
Fernando Zvietcovich and Kirill~V. Larin.
\newblock Wave-based optical coherence elastography: the 10-year perspective.
\newblock {\em Progress in Biomedical Engineering}, 4(1):012007, 2022.

\bibitem{leartprapun_2023}
Nichaluk Leartprapun and Steven~G. Adie.
\newblock Recent advances in optical elastography and emerging opportunities in
  the basic sciences and translational medicine [{Invited}].
\newblock {\em Biomedical Optics Express}, 14(1):208--248, 2023.

\bibitem{kirby_2017}
Mitchell~A. Kirby, Ivan Pelivanov, Shaozhen Song, Lukasz Ambrozinski, Soon~Joon
  Yoon, Liang Gao, David Li, Tueng~T. Shen, Ruikang~K. Wang, and Matthew
  O'Donnell.
\newblock Optical coherence elastography in ophthalmology.
\newblock {\em Journal of Biomedical Optics}, 22(12):121720, 2017.

\bibitem{lan_2023}
Gongpu Lan, Michael~D. Twa, Chengjin Song, JinPing Feng, Yanping Huang,
  Jingjiang Xu, Jia Qin, Lin An, and Xunbin Wei.
\newblock In vivo corneal elastography: {A} topical review of challenges and
  opportunities.
\newblock {\em Computational and Structural Biotechnology Journal},
  21:2664--2687, 2023.

\bibitem{low_2016}
Gavin Low, Scott~A. Kruse, and David~J. Lomas.
\newblock General review of magnetic resonance elastography.
\newblock {\em World Journal of Radiology}, 8(1):59--72, 2016.

\bibitem{sack_2023}
Ingolf Sack.
\newblock Magnetic resonance elastography from fundamental soft-tissue
  mechanics to diagnostic imaging.
\newblock {\em Nature Reviews Physics}, 5(1):25--42, January 2023.

\bibitem{marlevi_2020}
David Marlevi, Sharon~L. Mulvagh, Runqing Huang, J.~Kevin DeMarco, Hideki Ota,
  John Huston, Reidar Winter, Thanila~A. Macedo, Sahar~S. Abdelmoneim, Matilda
  Larsson, Patricia~A. Pellikka, and Matthew~W. Urban.
\newblock Combined spatiotemporal and frequency-dependent shear wave
  elastography enables detection of vulnerable carotid plaques as validated by
  {MRI}.
\newblock {\em Scientific Reports}, 10(1):403, 2020.

\bibitem{hiscox_2016}
Lucy~V. Hiscox, Curtis~L. Johnson, Eric Barnhill, Matt D.~J. McGarry, John
  Huston, Edwin J. R.~van Beek, John~M. Starr, and Neil Roberts.
\newblock Magnetic resonance elastography ({MRE}) of the human brain:
  technique, findings and clinical applications.
\newblock {\em Physics in Medicine \& Biology}, 61(24):R401, 2016.

\bibitem{shiina_2014}
Tsuyoshi Shiina.
\newblock Ultrasound elastography: {Development} of novel technologies and
  standardization.
\newblock {\em Japanese Journal of Applied Physics}, 53(7S):07KA02, 2014.

\bibitem{sigrist_2017}
Rosa~M.S. Sigrist, Joy Liau, Ahmed~El Kaffas, Maria~Cristina Chammas, and
  Juergen~K. Willmann.
\newblock Ultrasound {Elastography}: {Review} of {Techniques} and {Clinical}
  {Applications}.
\newblock {\em Theranostics}, 7(5):1303--1329, 2017.

\bibitem{doherty_2013}
J.~R. Doherty, G.~E. Trahey, K.~R. Nightingale, and M.~L. Palmeri.
\newblock Acoustic radiation force elasticity imaging in diagnostic ultrasound.
\newblock {\em IEEE Transactions on Ultrasonics, Ferroelectrics and Frequency
  Control}, 60(4):685--701, 2013.

\bibitem{bercoff_2004}
J.~Bercoff, M.~Tanter, and M.~Fink.
\newblock Supersonic shear imaging: a new technique for soft tissue elasticity
  mapping.
\newblock {\em IEEE Transactions on Ultrasonics, Ferroelectrics and Frequency
  Control}, 51(4):396--409, 2004.

\bibitem{deffieux_2008}
Thomas Deffieux.
\newblock {\em Palpation par force de radiation ultrasonore et échographie
  ultrarapide : {Applications} à la caractérisation tissulaire in vivo}.
\newblock Thèse de doctorat, Université Paris 7 - Paris Diderot, Institut
  Langevin, 2008.

\bibitem{kennedy_2014}
Brendan~F. Kennedy, Kelsey~M. Kennedy, and David~D. Sampson.
\newblock A {Review} of {Optical} {Coherence} {Elastography}: {Fundamentals},
  {Techniques} and {Prospects}.
\newblock {\em IEEE Journal of Selected Topics in Quantum Electronics},
  20(2):272--288, 2014.

\bibitem{li_2017}
Guo-Yang Li, Qiong He, Robert Mangan, Guoqiang Xu, Chi Mo, Jianwen Luo, Michel
  Destrade, and Yanping Cao.
\newblock Guided waves in pre-stressed hyperelastic plates and tubes:
  {Application} to the ultrasound elastography of thin-walled soft materials.
\newblock {\em Journal of the Mechanics and Physics of Solids}, 102:67--79,
  2017.

\bibitem{pelivanov_2019}
Ivan Pelivanov, Liang Gao, John Pitre, Mitchell~A. Kirby, Shaozhen Song, David
  Li, Tueng~T. Shen, Ruikang~K. Wang, and Matthew O'Donnell.
\newblock Does group velocity always reflect elastic modulus in shear wave
  elastography?
\newblock {\em Journal of Biomedical Optics}, 24(7):076003, 2019.

\bibitem{ramier_2019}
Antoine Ramier, Behrouz Tavakol, and Seok-Hyun Yun.
\newblock Measuring mechanical wave speed, dispersion, and viscoelastic modulus
  of the cornea using optical coherence elastography.
\newblock {\em Optics Express}, 27(12):16635, 2019.

\bibitem{bilston_2018}
Lynne~E. Bilston.
\newblock Soft tissue rheology and its implications for elastography:
  {Challenges} and opportunities.
\newblock {\em NMR in Biomedicine}, 31(10), 2018.

\bibitem{caenen_2022}
Annette Caenen, Mathieu Pernot, Kathryn~R Nightingale, Jens-Uwe Voigt,
  Hendrik~J Vos, Patrick Segers, and Jan D’hooge.
\newblock Assessing cardiac stiffness using ultrasound shear wave elastography.
\newblock {\em Physics in Medicine \& Biology}, 67(2):02TR01, 2022.

\bibitem{crutison_2022}
Joseph Crutison, Michael Sun, and Thomas~J. Royston.
\newblock The combined importance of finite dimensions, anisotropy, and
  pre-stress in acoustoelastography.
\newblock {\em The Journal of the Acoustical Society of America},
  151(4):2403--2413, 2022.

\bibitem{gennisson_2010}
Jean-Luc Gennisson, Thomas Deffieux, Emilie Macé, Gabriel Montaldo, Mathias
  Fink, and Mickaël Tanter.
\newblock Viscoelastic and {Anisotropic} {Mechanical} {Properties} of in vivo
  {Muscle} {Tissue} {Assessed} by {Supersonic} {Shear} {Imaging}.
\newblock {\em Ultrasound in Medicine \& Biology}, 36(5):789--801, 2010.

\bibitem{yasar_2013}
Temel~K. Yasar, Thomas~J. Royston, and Richard~L. Magin.
\newblock Wideband {MR} elastography for viscoelasticity model identification.
\newblock {\em Magnetic Resonance in Medicine}, 70(2):479--489, 2013.

\bibitem{brum_2014}
J.~Brum, M.~Bernal, J.~L. Gennisson, and M.~Tanter.
\newblock In vivo evaluation of the elastic anisotropy of the human {Achilles}
  tendon using shear wave dispersion analysis.
\newblock {\em Physics in Medicine \& Biology}, 59(3):505, January 2014.

\bibitem{zampini_2021}
Marco~Andrea Zampini, Martina Guidetti, Thomas~J. Royston, and Dieter Klatt.
\newblock Measuring viscoelastic parameters in {Magnetic} {Resonance}
  {Elastography}: a comparison at high and low magnetic field intensity.
\newblock {\em Journal of the Mechanical Behavior of Biomedical Materials},
  120:104587, 2021.

\bibitem{juge_2023}
Lauriane Jugé, Patrick Foley, Alice Hatt, Jade Yeung, and Lynne~E. Bilston.
\newblock Ex vivo bovine liver nonlinear viscoelastic properties: {MR}
  elastography and rheological measurements.
\newblock {\em Journal of the Mechanical Behavior of Biomedical Materials},
  138:105638, 2023.

\bibitem{sharma_2023}
Ananya Sharma, Sai~Geetha Marapureddy, Abhijit Paul, Sapna~R. Bisht, Manik
  Kakkar, Prachi Thareja, and Karla~P. Mercado-Shekhar.
\newblock Characterizing {Viscoelastic} {Polyvinyl} {Alcohol} {Phantoms} for
  {Ultrasound} {Elastography}.
\newblock {\em Ultrasound in Medicine and Biology}, 49(2):497--511, 2023.

\bibitem{couade_2010}
Mathieu Couade, Mathieu Pernot, Claire Prada, Emmanuel Messas, Joseph Emmerich,
  Patrick Bruneval, Aline Criton, Mathias Fink, and Mickael Tanter.
\newblock Quantitative {Assessment} of {Arterial} {Wall} {Biomechanical}
  {Properties} {Using} {Shear} {Wave} {Imaging}.
\newblock {\em Ultrasound in Medicine \& Biology}, 36(10):1662--1676, 2010.

\bibitem{astaneh_2017}
Ali~Vaziri Astaneh, Matthew~W. Urban, Wilkins Aquino, James~F. Greenleaf, and
  Murthy~N. Guddati.
\newblock Arterial waveguide model for shear wave elastography: implementation
  and in vitro validation.
\newblock {\em Physics in Medicine \& Biology}, 62(13):5473, 2017.

\bibitem{maksuti_2017}
Elira Maksuti, Fabiano Bini, Stefano Fiorentini, Giulia Blasi, Matthew~W.
  Urban, Franco Marinozzi, and Matilda Larsson.
\newblock Influence of wall thickness and diameter on arterial shear wave
  elastography: a phantom and finite element study.
\newblock {\em Physics in Medicine \& Biology}, 62(7):2694, 2017.

\bibitem{nenadic_2011}
Ivan~Z. Nenadic, Matthew~W. Urban, Scott~A. Mitchell, and James~F. Greenleaf.
\newblock Lamb wave dispersion ultrasound vibrometry ({LDUV}) method for
  quantifying mechanical properties of viscoelastic solids.
\newblock {\em Physics in Medicine \& Biology}, 56(7):2245, 2011.

\bibitem{nguyen_2011}
{Thu-Mai Nguyen}, M.~Couade, J.~Bercoff, and M.~Tanter.
\newblock Assessment of viscous and elastic properties of sub-wavelength
  layered soft tissues using shear wave spectroscopy: {Theoretical} framework
  and in vitro experimental validation.
\newblock {\em IEEE Transactions on Ultrasonics, Ferroelectrics and Frequency
  Control}, 58(11):2305--2315, 2011.

\bibitem{catheline_2003}
S.~Catheline, J.-L. Gennisson, and M.~Fink.
\newblock Measurement of elastic nonlinearity of soft solid with transient
  elastography.
\newblock {\em The Journal of the Acoustical Society of America},
  114(6):3087--3091, December 2003.

\bibitem{gennisson_2007}
J.-L. Gennisson, M.~Rénier, S.~Catheline, C.~Barrière, J.~Bercoff, M.~Tanter,
  and M.~Fink.
\newblock Acoustoelasticity in soft solids: {Assessment} of the nonlinear shear
  modulus with the acoustic radiation force.
\newblock {\em The Journal of the Acoustical Society of America},
  122(6):3211--3219, 2007.

\bibitem{biot_1940}
Maurice~A. Biot.
\newblock The {Influence} of {Initial} {Stress} on {Elastic} {Waves}.
\newblock {\em Journal of Applied Physics}, 11(8):522--530, 1940.

\bibitem{toupin_1961}
R.~A. Toupin and B.~Bernstein.
\newblock Sound {Waves} in {Deformed} {Perfectly} {Elastic} {Materials}.
  {Acoustoelastic} {Effect}.
\newblock {\em The Journal of the Acoustical Society of America},
  33(2):216--225, 1961.

\bibitem{ogden_1997}
R.~W. Ogden.
\newblock {\em Non-linear elastic deformations}.
\newblock Dover Publications, Mineola, N.Y, 1997.

\bibitem{destrade_2007}
Michel Destrade and G.~Saccomandi.
\newblock {\em Waves in nonlinear pre-stressed materials}.
\newblock Number 495 in {CISM} {Courses} and {Lectures}. Udine, Italy, springer
  wien new york edition, 2007.

\bibitem{zhang_2023}
Zhaoyi Zhang, Guo-Yang Li, Yuxuan Jiang, Yang Zheng, Artur~L. Gower, Michel
  Destrade, and Yanping Cao.
\newblock Noninvasive measurement of local stress inside soft materials with
  programmed shear waves.
\newblock {\em Science Advances}, 9(10):eadd4082, 2023.
\newblock Publisher: American Association for the Advancement of Science.

\bibitem{li_cao_2017}
Guo-Yang Li and Yanping Cao.
\newblock Mechanics of ultrasound elastography.
\newblock {\em Proceedings of the Royal Society A: Mathematical, Physical and
  Engineering Sciences}, 473(2199):20160841, 2017.

\bibitem{laurent_2020}
Jérôme Laurent, Daniel Royer, and Claire Prada.
\newblock In-plane backward and zero group velocity guided modes in rigid and
  soft strips.
\newblock {\em The Journal of the Acoustical Society of America},
  147(2):1302--1310, 2020.

\bibitem{lanoy_2020}
Maxime Lanoy, Fabrice Lemoult, Antonin Eddi, and Claire Prada.
\newblock Dirac cones and chiral selection of elastic waves in a soft strip.
\newblock {\em Proceedings of the National Academy of Sciences},
  117(48):30186--30190, 2020.

\bibitem{delory_2022}
Alexandre Delory, Fabrice Lemoult, Maxime Lanoy, Antonin Eddi, and Claire
  Prada.
\newblock Soft elastomers: {A} playground for guided waves.
\newblock {\em The Journal of the Acoustical Society of America},
  151(5):3343--3358, 2022.

\bibitem{delory_2023b}
Alexandre Delory, Daniel~A. Kiefer, Maxime Lanoy, Antonin Eddi, Claire Prada,
  and Fabrice Lemoult.
\newblock Elastodynamics of a soft strip subject to a large deformation, 2023.

\bibitem{delory_2023a}
Alexandre Delory, Fabrice Lemoult, Antonin Eddi, and Claire Prada.
\newblock Guided elastic waves in a highly-stretched soft plate.
\newblock {\em Extreme Mechanics Letters}, page 102018, 2023.

\bibitem{kiefer_2019}
Daniel~A. Kiefer, Michael Ponschab, Stefan~J. Rupitsch, and Michael Mayle.
\newblock {Calculating the full leaky Lamb wave spectrum with exact fluid
  interaction}.
\newblock {\em The Journal of the Acoustical Society of America},
  145(6):3341--3350, 06 2019.

\bibitem{liao_2021}
Zisheng Liao, Jie Yang, Mokarram Hossain, Gregory Chagnon, Lin Jing, and Xiaohu
  Yao.
\newblock On the stress recovery behaviour of {Ecoflex} silicone rubbers.
\newblock {\em International Journal of Mechanical Sciences}, 206:106624, 2021.

\bibitem{liu_2014}
Yifei Liu, Temel~K. Yasar, and Thomas~J. Royston.
\newblock Ultra wideband (0.5–16 {kHz}) {MR} elastography for robust shear
  viscoelasticity model identification.
\newblock {\em Physics in Medicine \& Biology}, 59(24):7717, 2014.

\bibitem{kearney_2015}
Steven~P. Kearney, Altaf Khan, Zoujun Dai, and Thomas~J. Royston.
\newblock Dynamic viscoelastic models of human skin using optical elastography.
\newblock {\em Physics in Medicine \& Biology}, 60(17):6975, 2015.

\bibitem{royer2022elastic}
Daniel Royer and Tony Valier-Brasier.
\newblock {\em Elastic Waves in Solids, Volume 1: Propagation}.
\newblock John Wiley \& Sons, 2022.

\bibitem{auld1973acoustic}
B.~A. Auld.
\newblock {\em Acoustic Fields and Waves in Solids}.
\newblock John Wiley \& Sons Inc., 1973.

\bibitem{parker_2019}
K.~J. Parker, T.~Szabo, and S.~Holm.
\newblock Towards a consensus on rheological models for elastography in soft
  tissues.
\newblock {\em Physics in Medicine \& Biology}, 64(21):215012, 2019.

\bibitem{baranger_2023}
Jerome Baranger, Olivier Villemain, Guillaume Goudot, Alexandre Dizeux,
  Heiva~Le Blay, Tristan Mirault, Emmanuel Messas, Mathieu Pernot, and Mickael
  Tanter.
\newblock The fundamental mechanisms of the korotkoff sounds generation.
\newblock {\em Science Advances}, 9(40):eadi4252, 2023.

\bibitem{rivlin1948large}
Ronald~S Rivlin.
\newblock Large elastic deformations of isotropic materials iv. further
  developments of the general theory.
\newblock {\em Philosophical transactions of the royal society of London.
  Series A, Mathematical and physical sciences}, 241(835):379--397, 1948.

\bibitem{wex_2015}
Cora Wex, Susann Arndt, Anke Stoll, Christiane Bruns, and Yuliya Kupriyanova.
\newblock Isotropic incompressible hyperelastic models for modelling the
  mechanical behaviour of biological tissues: a review.
\newblock {\em Biomedical Engineering}, 60(6), 2015.

\bibitem{chagnon_2015}
G.~Chagnon, M.~Rebouah, and D.~Favier.
\newblock Hyperelastic energy densities for soft biological tissues: A review.
\newblock {\em Journal of Elasticity}, 120:129--160, 2015.

\bibitem{balzani_2006}
D.~Balzani, P.~Neff, J.~Schröder, and G.A. Holzapfel.
\newblock A polyconvex framework for soft biological tissues. {Adjustment} to
  experimental data.
\newblock {\em International Journal of Solids and Structures},
  43(20):6052--6070, 2006.

\bibitem{peyraut_2010}
F.~Peyraut, Z.-Q. Feng, N.~Labed, and C.~Renaud.
\newblock A closed form solution for the uniaxial tension test of biological
  soft tissues.
\newblock {\em International Journal of Non-Linear Mechanics}, 45(5):535--541,
  2010.

\bibitem{mukherjee_2022}
Soumya Mukherjee, Michel Destrade, and Artur~L. Gower.
\newblock Representing the stress and strain energy of elastic solids with
  initial stress and transverse texture anisotropy.
\newblock {\em Proceedings of the Royal Society A: Mathematical, Physical and
  Engineering Sciences}, 478(2266):20220255, 2022.

\end{thebibliography}

\end{document}